\documentclass[]{emulateapj}

\def\kms{km s$^{-1}$}
\def\msun{M$_{\odot}$}
\def\rsun{R$_{\odot}$}

\begin{document}

\title{A 12 minute Orbital Period Detached White Dwarf Eclipsing
Binary\altaffilmark{*}}

\author{Warren R.\ Brown$^1$,
	Mukremin Kilic$^{1,}$\altaffilmark{**},
	J.\ J.\ Hermes$^2$,
	Carlos Allende Prieto$^{3,4}$,
	Scott J.\ Kenyon$^1$,
	and
	D.\ E.\ Winget$^2$,
	}

\affil{ $^1$Smithsonian Astrophysical Observatory, 60 Garden St, Cambridge, MA 02138, USA\\
	$^2$Department of Astronomy, University of Texas at Austin, RLM 16.236, Austin, TX 78712, USA\\
	$^3$Instituto de Astrof\'{\i}sica de Canarias, E-38205, La Laguna, Tenerife, Spain\\
	$^4$Departamento de Astrof\'{\i}sica, Universidad de La Laguna, E-38205 La Laguna, Tenerife, Spain
	}

\email{wbrown@cfa.harvard.edu, mkilic@cfa.harvard.edu}


\altaffiltext{*}{Based on observations obtained at the MMT Observatory, a joint 
facility of the Smithsonian Institution and the University of Arizona, and on 
observations obtained at The McDonald Observatory of The University of Texas at 
Austin.}

\altaffiltext{**}{\em Spitzer Fellow}

\shorttitle{ A 12 Minute Orbital Period Detached Eclipsing Binary }
\shortauthors{Brown et al.}

\begin{abstract}
	We have discovered a detached pair of white dwarfs (WDs) with a 12.75 min
orbital period and a 1,315 \kms\ radial velocity amplitude.  We measure the full
orbital parameters of the system using its light curve, which shows ellipsoidal
variations, Doppler boosting, and primary and secondary eclipses.  The primary is a
0.25 \msun\ tidally distorted helium WD, only the second tidally distorted WD known.  
The unseen secondary is a 0.55 \msun\ carbon-oxygen WD.  The two WDs will come into 
contact in 0.9 Myr due to loss of energy and angular momentum via gravitational wave 
radiation.  Upon contact the systems may merge yielding a rapidly spinning massive WD, 
form a stable interacting binary, or possibly explode as an underluminous supernova 
type Ia.  The system currently has a gravitational wave strain of $10^{-22}$,
about 10,000 times larger than the Hulse-Taylor pulsar;  this system would be
detected by the proposed LISA gravitational wave mission in the first week of
operation.  This system's rapid change in orbital period will provide a fundamental
test of general relativity.

\end{abstract}


\slugcomment{Accepted to ApJ Letters}

\section{INTRODUCTION}

	In Einstein's general theory of relativity, close pairs of stars produce
gravitational waves, ripples in the curvature of space-time
\citep{einstein16,einstein18}.  Observations of the Hulse-Taylor binary pulsar PSR
B1913+16 confirm the slow decay of the orbit predicted by Einstein's theory and
provide indirect evidence for gravitational waves \citep{hulse75,weisberg10}.  
On-going and proposed instruments seek to detect gravitational waves directly
\citep{jafry94,hobbs09,abbott10}.  The strongest known gravitational wave sources
are compact binaries containing neutron stars and WDs \citep[e.g.,][]{nelemans09}.  
There are presently four known binaries with orbital periods less than 15 minutes --
three AM CVn stars \citep{haberl95,israel99,warner02} and one low mass X-ray binary
\citep{stella87} -- but all binaries with $<$1 hr orbital periods, with one
exception \citep{kilic11b}, are interacting systems that complicate direct tests of
Einstein's theory.  \citet{marsh05} show that the expected period change due to
gravitational wave radiation can be modified by both induction-driven angular
momentum interchange in magnetic WD systems and by mass-transfer in accreting
systems.  Moreover, there are different explanations for both the nature of the
shortest-period AM CVn systems and their observed period changes
\citep{strohmayer04,dantona06,deloye06,roelofs10}.

	Here, we report the discovery of a detached, 765 sec orbital period binary,
SDSS J065133.33+284423.3 (hereafter J0651).  The discovery comes from our on-going
ELM Survey, a targeted spectroscopic survey for extremely low mass $<$0.25\msun\ WDs
\citep{brown10c,kilic10}.  The ELM Survey demonstrates that extremely low mass WDs
are found in compact binaries, and that the majority of these binaries have $<$10
Gyr merger times due to gravitational wave radiation \citep{kilic11a}. Depending on
the poorly constrained physics of the merger process, these merging WD systems may
produce single helium-rich sdO stars, stable mass-transfer AM CVn binaries, or
possibly underluminous supernova \citep[as discussed in][]{kilic10}.  
The observed merger rate of the low mass WD binaries is about 1\% of the Type Ia
supernovae rate -- comparable to the rate of underluminous supernovae
\citep{brown11a}.  The ELM Survey is also responsible for finding the first tidally
distorted WD \citep{kilic11b}.  J0651 is the most extreme system discovered to date
and shows that the ELM Survey may open a new window on gravitational wave sources.

	In J0651, the orientation of the binary allows us to observe eclipses of
each star by the other, leading to accurate measurement of the orbital parameters,
masses, and WD radii.  There is no evidence for mass transfer.  These results
suggest that J0651 is the cleanest known strong gravitational wave source, with an
orbital decay time of less than 1 Myr.  Future observations will allow us to measure
the change in the orbital period.  We hope to one day compare this change with
direct measurements of gravitational waves and provide an unprecedented test of
general relativity.

	~

\begin{figure}		
 \plotone{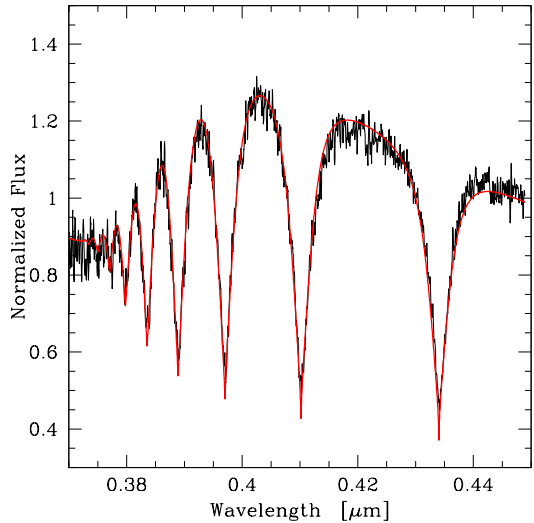}
 \caption{ \label{fig:spec}
	The composite observed spectrum of J0651, constructed by shifting the
individual spectra to rest-frame and then summing them together to create an
effective 60.5 min exposure, compared to the best-fit stellar atmosphere (solid red
line)  for a 0.25 \msun\ WD.  The broad hydrogen Balmer absorption lines visually
indicate J0651 is an unusually low mass WD.}
	\end{figure}

\section{DATA AND ANALYSIS}

	We discovered J0651 on 2011 March 2 as part of our targeted spectroscopic 
survey for extremely low mass WDs using the 6.5m MMT telescope.  We used the MMT 
Blue Channel spectrograph \citep{schmidt89} to obtain 1 \AA\ resolution spectra in 
the wavelength range 3600 \AA\ - 4500 \AA .  We recognized J0651 as a low mass WD 
due to its pressure-broadened hydrogen Balmer lines, seen in Figure \ref{fig:spec}.

	Spectral fits to hydrogen atmosphere WD models \citep{koester08} yield a 
surface gravity of $\log{g}=6.79\pm0.04$ dex ($g$ in cm s$^{-2}$) and an effective 
temperature of $T_{\rm eff}=16,400\pm300$ K.  The Sloan Digital Sky Survey 
\citep[SDSS,][]{aihara11} spectrum of this object, on the other hand, yields a 
systematically larger $\log{g}=6.97\pm0.05$ dex and $T_{\rm eff}=17,700\pm250$ K 
(S.~J.\ Kleinman 2011, private communication) because the SDSS exposure spans 3.5 
orbital periods and thus measures artificially broadened Balmer lines.  Our own 6-10 
min exposures of J0651 show similarly over-estimated surface gravities and 
temperatures.  A correct understanding of J0651 requires time-resolved spectroscopy 
only possible with a larger telescope like the 6.5m MMT.

\begin{figure}		
 \plotone{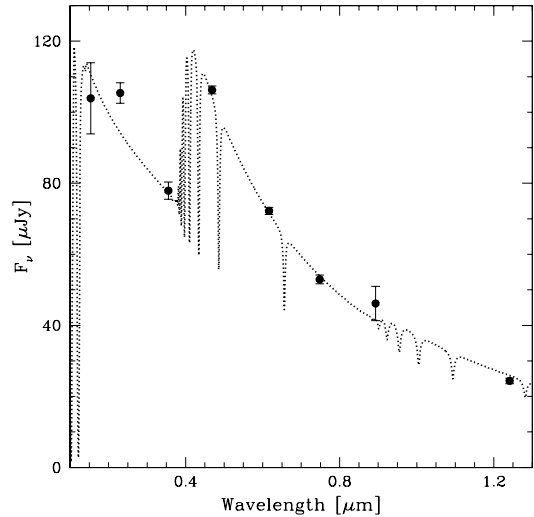}
 \caption{ \label{fig:sed}
	Best-fit WD model atmosphere (dotted line) compared to broadband photometry
(dots).  The ultraviolet, optical, and near-infrared measurements support our
spectroscopic fit.}
	\end{figure}

	Figure \ref{fig:sed} compares our best-fit model atmosphere to dereddened 
($E(B-V)$ = 0.0706 mag) ultraviolet Galaxy Evolution Explorer (GALEX) photometry 
\citep{martin05}, optical SDSS photometry, and our own near-infrared 
$J=19.599\pm0.029$ mag measurement obtained with SWIRC \citep{brown08c} on the MMT.  
The broadband photometry supports our spectroscopic effective temperature and 
surface gravity measurement with one possible exception.  The GALEX NUV point shows 
a 2-$\sigma$ discrepancy with the model spectrum convolved with the filter bandpass, 
however the photometric error does not include the significant uncertainty in the 
extinction correction for GALEX filters \citep{wyder05} and so we consider the 
discrepancy suspect.  Comparison with WD evolutionary tracks \citep{panei07} 
indicates that J0651 is consistent with an extremely low mass 0.25 \msun\ 
helium-core WD.

\begin{figure}		
 \plotone{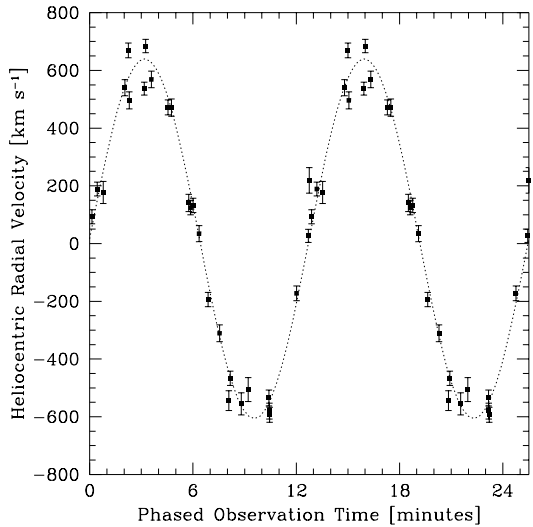}
 \caption{ \label{fig:rv}
	Radial velocity observations phased to the 765 sec orbital period.  The 
best-fit orbit (dotted line) has a 1314.6 \kms\ velocity amplitude, which is 0.44\% 
the speed of light.}
	\end{figure}

	We discovered that J0651 is a compact binary system when back-to-back 
spectra separated by 6 min showed a $\simeq$1300 \kms\ change in radial velocity 
(Figure \ref{fig:rv}).  Despite the faintness of the $g=19.1$ WD, the 
light-gathering power of the 6.5m MMT telescope allowed us to reduce our exposure 
times to 2-2.5 minutes and resolve the orbital period.  The data are presented in 
Table \ref{tab:dat}.  The observed velocity amplitude is an underestimate, however, 
because 1) our exposures span 18\% of orbital phase and 2) the radial velocity curve 
is sinusoidal, not linear.  By integrating a sine curve at the phase covered by our 
exposures, we determine that the velocity amplitude correction is 5.5\%.  The 
observed velocity amplitude is also underestimated if the WD is tidally distorted 
and its center-of-light differs from its center-of-mass.  We do not correct for this 
effect given that the velocity correction is less than 1\% -- comparable to our 
measurement uncertainty -- for the observed oblateness.

	The best-fit orbital parameters to the radial velocities are:  period 
$P=765.4\pm7.9$ sec, corrected velocity semi-amplitude $K=657.3\pm2.4$ \kms , and 
systemic velocity $\gamma=16.6\pm0.6$ \kms .  J0651 is very likely a Galactic disk 
object based on its small proper motion $(\mu_\alpha,\mu_\delta)=(-3.6,-1.2)$ mas 
yr$^{-1}$ \citep{munn04} and small systemic velocity.  J0651's location 220 pc above 
the plane (for a distance of 1 kpc, see below) is also consistent with a disk 
object.

\begin{figure}		
 \plotone{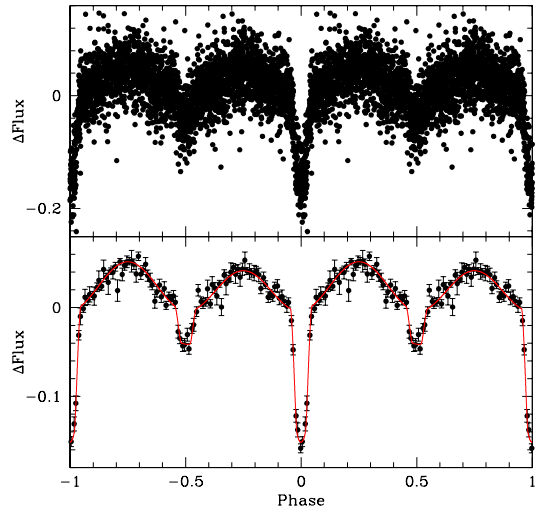}
 \caption{ \label{fig:lc}
	J0651 light curve.  The upper panel plots the observed photometry vs.\ 
orbital phase, while the lower panel compares the binned data to the best-fit 
model (solid red line).  The data reveals three dramatic features: a sinusoidal 
pattern due to ellipsoidal variations from the tidally distorted WD, an asymmetric 
peak in light due to Doppler boosting, and periodic dips in light from the eclipses 
of the primary (at phase 0) and secondary (at phase 0.5) WDs. }
	\end{figure}

	We obtained time-series optical photometry of J0651 using the McDonald 2.1m
Otto Struve Telescope using the Argos frame transfer camera \citep{nather04}.  
Images were obtained with a {\it BG40} Schott glass filter. Our set of 3,038 10-
and 15-second exposures come from five different nights in 2011 April covering a
total time baseline of 12.1 days.  The observed light curve, phased to the best-fit
period, is plotted in Figure \ref{fig:lc} and shows three significant features: a
sinusoidal pattern due to ellipsoidal variations from the tidally distorted WD, an
asymmetric peak in light due to relativistic beaming (so-called Doppler boosting),
and periodic dips in light from the eclipses of the primary and secondary WDs.

\begin{deluxetable}{cr}
\tablecolumns{2}
\tablewidth{0pt}
\tablecaption{Radial Velocity Measurements\label{tab:dat}}
\tablehead{
	\colhead{HJD+2455600}   & \colhead{$v_{helio}$}\\
	(days){\dag}    & (km s$^{-1}$)
}
	\startdata
22.76724617 &  471.0 $\pm$ 30.2 \\
22.76919045 & -310.7 $\pm$ 28.9 \\
22.77118102 & -534.9 $\pm$ 27.8 \\
22.77311373 &  189.3 $\pm$ 23.9 \\
22.77506958 &  684.5 $\pm$ 23.1 \\
22.77697914 &  132.2 $\pm$ 25.0 \\
22.77892341 & -555.3 $\pm$ 38.4 \\
22.78221017 &  177.1 $\pm$ 38.3 \\
22.78415444 &  568.7 $\pm$ 29.0 \\
22.78607558 &   34.5 $\pm$ 27.7 \\
22.78806615 & -505.9 $\pm$ 41.3 \\
22.79001043 & -172.5 $\pm$ 25.4 \\
22.79194313 &  540.2 $\pm$ 28.0 \\
23.66752771 &   92.8 $\pm$ 24.4 \\
23.66963400 &  537.0 $\pm$ 22.3 \\
23.67142783 &  140.8 $\pm$ 29.7 \\
23.67304806 & -544.2 $\pm$ 34.5 \\
23.67469144 & -593.0 $\pm$ 26.1 \\
23.67628852 &  218.9 $\pm$ 44.2 \\
23.67788561 &  496.0 $\pm$ 29.6 \\
23.68036224 &  124.7 $\pm$ 25.4 \\
23.68195933 & -466.8 $\pm$ 25.0 \\
23.68353327 & -580.2 $\pm$ 27.4 \\
23.68510721 &   26.8 $\pm$ 22.8 \\
23.68670429 &  669.3 $\pm$ 25.5 \\
23.68830137 &  471.9 $\pm$ 26.0 \\
23.68992161 & -193.7 $\pm$ 24.5 \\
	\enddata
 \tablecomments{\dag Based on UTC.}
\end{deluxetable}

	We model the light curve of J0651 using JKTEBOP \citep{southworth04} and 
verify our results with PHOEBE \citep{prsa05}. JKTEBOP\footnote{
	JKTEBOP models the projection of each star as a biaxial ellipsoid and 
calculates a light curve by numerical integration of concentric circles over each 
star. A by-product of this calculation is an estimate of the oblateness of the 
primary WD.  The 3\% oblateness of J0651 falls within JKTEBOP's 4\% oblateness limit 
for accurate light curve analysis.}
	and PHOEBE are based on the Eclipsing Binary Orbit Program \citep{popper81} and 
the \citet{wilson71} codes, respectively.  We first remove the 0.5\% Doppler boosting 
(relativistic beaming)  signal, however, because neither code models Doppler boosting.  
Doppler boosting has been seen in only a handful of systems 
\citep{vankerkwijk10,mazeh10,shporer10,bloemen11}; its asymmetric contribution to the 
J0651 light curve confirms the 765 sec orbital period.  We then fit the 5\% amplitude 
ellipsoidal variations and the 15\% primary eclipse and 4\% secondary eclipse.  
Reflection effects due to the heating of each WD by its companion are weak, 
0.3\%$\pm$0.4\%.

	The light curve yields a much more precise measurement of orbital period,
$P=765.2062\pm0.003$ sec. The ellipsoidal variations are due to the changing
projected area of the distorted primary WD, hence they strongly depend on the
inclination angle of the binary system. J0651's ellipsoidal variations and eclipses
constrain\footnote{Errors are estimated from 10,000 Monte-Carlo simulations with
JKTEBOP as described in \citet{southworth05}.} the orbital inclination to
$i=86.9^{+1.6}_{-1.0}$ deg.

	Eclipses also provide a precise measurement of the WD radii.  The 0.25 
\msun\ primary WD has an observed radius of 0.0353$\pm$0.0004 \rsun\ that differs by 
5\% from the 0.0337 \rsun\ radius predicted by helium WD models \citep{panei07}. 
Going in the other direction, the models predict that a WD with the observed radius 
and with mass 0.24 \msun\ has $\log{g}=6.71$ dex and $T_{\rm eff}=16,000$ K, 
consistent with our spectroscopic observations.  Given the uncertainties, we 
consider the observed and predicted radii in excellent agreement.

	The unseen secondary has a mass of 0.55 \msun\ and an observed radius of 
0.0132$\pm$0.0003 \rsun\ typical for a carbon-oxygen WD.  The eclipse depths 
indicate the secondary is $3.7\% \pm 0.3\%$ as bright as the primary.  Adopting 
$M_g=8.9\pm0.1$ mag as the absolute magnitude of the 0.25 \msun\ primary 
\citep{panei07}, the secondary has $M_g=12.5$ mag.  A carbon-oxygen WD of this 
luminosity has $T_{\rm eff}=9000$ K, $\log{g}=7.9$ dex, a cooling age of 700 Myr, 
and a radius of 0.0137 \rsun\ (P.-E.\ Tremblay 2011, private communication).  The 
observed secondary radius differs by 4\% from the predicted radius.  There are only 
two other model-independent mass and radius determinations of helium and 
carbon-oxygen WDs \citep{steinfadt10,parsons11}.  Our results thus provide important 
constraints and reveal an overall agreement with current WD models.

\section{DISCUSSION}

	Having accurately measured J0651's binary parameters we now calculate the 
gravitational wave emission predicted by Einstein's general theory of relativity.  
J0651, with a de-reddened apparent magnitude $g_0=18.84\pm0.01$ and thus a 
$1.0\pm0.1$ kpc distance from the Sun, has a predicted gravitational wave strain 
\citep[e.g.,][]{roelofs07a} of $h=1.2\times10^{-22}$.  J0651 is among the strongest 
known gravitational wave sources and, more importantly, has an orbital frequency 
that places it well above the expected gravitational wave foreground 
\citep{nelemans09}. The proposed ESA/NASA Laser Interferometer Space Antenna ({\it 
LISA}) mission has its peak sensitivity at frequencies corresponding to $\simeq$5 
min orbital periods, and it should detect a strain of $10^{-22}$ at these 
frequencies with a signal-to-noise of $\simeq$100 in one year \citep{roelofs07a}.  
In other words, J0651 is a verification source that {\it LISA}, if built, would 
detect in the first week of operation.

	We also predict that J0651's orbital period is shrinking by $2.7 \times 
10^{-4}$ sec per year due to gravitational wave radiation.  The expected change in 
period adds up to a 5.5 sec change in time-of-eclipse in one year.  When we measure 
this change we expect to provide yet another fundamental test of general relativity 
and the existence of gravitational waves.

	The absence of mass transfer in J0651 is perhaps surprising given how 
quickly it will merge.  In all known binaries with periods comparable to J0651 
\citep[e.g.,][]{nelemans09} one star fills its Roche lobe and transfers mass 
to its companion.  Our data show that the J0651 primary has a Roche lobe radius 1.5 
times its current radius.  Under the assumption of energy and angular momentum loss 
due to gravitational wave radiation, the primary WD will reach its Roche lobe radius 
at an orbital period of 6.5 minutes in 0.9 Myr.

	What happens when the primary WD fills its Roche lobe and begins mass
transfer in 0.9 Myr is an open theoretical question.  The stability of the mass
transfer phase depends in part on the known binary mass ratio $q=0.45$ and in part
on the unknown tidal synchronization and entropy of the primary WD \citep{marsh04}.  
If mass transfer is stable, J0651 will become an AM CVn system like HM Cancri
\citep{roelofs10}.  Recent WD mass transfer models, on the other hand, predict that
mass transfer will cause the pair of WDs to quickly coalesce and merge
\citep{dan11}.  A merger will likely produce a rapidly spinning massive WD, but an
underluminous supernova explosion is also a possibility.  Underluminous supernovae,
such as SN 2005E, are rare types of supernovae that are 10-100 times less luminous
than a normal supernova Type Ia and have only $\sim$0.25 \msun\ worth of ejecta
\citep{perets10a}.  The spectrum and light curve of SN 2005E, for example, can be
explained by the detonation of a 0.2 \msun\ helium layer on a 0.45 \msun\ WD
\citep{waldman10} -- parameters very similar to the J0651 system.  Depending on the
nature of the mass transfer, there may be other mechanisms by which the system could
detonate \citep{bildsten07,guillochon10}.  Completing our survey for extremely low
mass WD systems such as J0651 will allow us to measure the space density and merger
rate of these systems and compare with the rate of underluminous supernovae.

\section{CONCLUSION}

	The eclipsing detached WD binary system J0651 presents us with a remarkable 
laboratory.  We can use its eclipses to measure WD masses and radii for detailed 
tests of WD models.  We can use its changing binary orbital period and gravitational 
wave strain to test for the existence of gravitational waves predicted by general 
relativity.  We can use its very existence to constrain the space density and merger 
rate of low mass WD binaries and links to underluminous supernovae.  In the future, 
we plan to use multi-passband photometry to directly measure the nature of the 
secondary WD and to detect the change in orbital period predicted by general 
relativity.

\acknowledgments
	We thank John Kuehne for his assistance with the observations.  This work 
was supported in part by the Smithsonian Institution and by NASA through the
{\em Spitzer Space Telescope} Fellowship Program and in part by the National
Science Foundation under grant AST-0909107 and the Norman Hackerman Advanced 
Research Program under grant 003658-0252-2009.

\noindent {\it Facilities:} {MMT (Blue Channel Spectrograph, SWIRC)},
{2.1m Otto Struve Telescope (Argos)}


\clearpage


\end{document}